# Rollable Magnetoelectric Energy Harvester as Wireless IoT Sensor

Sujoy Kumar Ghosh,[†,‡] Krittish Roy,[†] Hari Krishna Mishra,[∥] Manas Ranjan Sahoo,[⊥] Biswajit Mahanty, [†,§] Prakash Nath Vishwakarma[⊥] and Dipankar Mandal[†,∥,*]

[†]Organic Nano-Piezoelectric Device Laboratory, Department of Physics
Jadavpur University, Kolkata 700032, India

[∥]Institute of Nano Science and Technology, Phase-10, Sector-64, Mohali, 160062, India

[§]Department of Electronics and Communication Engineering,
Saroj Mohan Institute of Technology, Hooghly 712512, India

[⊥]Department of Physics and Astronomy, National Institute of Technology,
Rourkela, Odisha 769008, India

[‡] This author currently working in Ulsan National Institute of Science and Technology (UNIST), South Korea.

*E-mail: dipankar@phys.jdvu.ac.in, dmandal@inst.ac.in. Tel.: +91-172-2210075. Fax: +91-172-2211074

**Abstract**

Perhaps the most abundant form of waste energy in our surrounding is the parasitic magnetic noise arising from electrical power transmission system. In this work, a flexible and rollable magneto-mechano-electric nanogenerator (MMENG) based wireless IoT sensor has been demonstrated in order to capture and utilize the magnetic noise. Free standing magnetoelectric composites are fabricated by combining magnetostrictive nickel ferrite ($NiFe_2O_4$) nanoparticles and piezoelectric polyvinylidene-co-trifluoroethylene (P(VDF–TrFE)) polymer. The magnetoelctric 0-3 type nanocomposites possess maximum magnetoelectric voltage co-efficient ($\alpha$) of 11.43 mV/cm-Oe. Even, without magnetic bias field 99 % of the maximum $\alpha$ value is observed due to self-bias effect. As a result, the MMENG generates peak-to-peak open circuit voltage of 1.4 V, output power density of 0.05 µW/cm$^3$ and successfully operates commercial capacitor under the weak ($\sim 1.7\times 10^{-3}$ T) and low frequency ($\sim$ 50 Hz) stray magnetic field arising from the power cable of home appliances such as, electric kettle. Finally, the harvested electrical signal has been wirelessly transmitted to a smart phone in order to demonstrate the possibility of position monitoring system construction. This cost effective and easy to integrate approach with tailored size and shape of device configuration is expected to be explored in next-generation self-powered IoT sensors including implantable biomedical devices and human health monitoring sensory systems.

1. Introduction

Energy harvesting is the process to convert the wasted environmental energy into utilizable electrical energy. Therefore, energy derived from ambient energy resources such as, solar, thermal, wind, vibration, and magnetic field is an area of focus for current and next generation autonomous electronic devices and wireless sensor networks [1,2]. In our everyday life, we are basically surrounded by 50/60 Hz parasitic magnetic noise arising from electrical power cables, electronic systems, subways and so on which indicates that magnetic field is a ubiquitous energy source for harvesting [3–5]. According to Ampere's law, any current carrying wires connected to the home appliances (such as, refrigeration units, lights, heating/cooling appliances, displays, and so on) can generate magnetic field which is of very low amplitude (~ mT) and frequency. Thus, it is a challenging task to harvest this low amplitude and frequency magnetic field in order to build a sustainable electricity source [4,5]. Conventionally, magnetic energy was harvested using electromagnetic devices using coils and magnets (Faraday's induction law) which have limitations related to frequency, size and efficiency [4,5]. Those limitations can be overcome through the designs of newly discovered magnetoelectric (ME) composite materials [5,6]. In the ME effect, the magnetic field induced magnetostrictive strain transferred to the piezoelectric layer across interface generates the electric polarization due to piezoelectric effect [5]. Therefore, the ME materials generate electrical voltage when subject to a low magnetic field. Since the ME effect is related to the interface condition, therefore, strong interfacial attachment between piezoelectric and magnetostrictive components is necessary for ME coupling and eventually ME co-efficient [7]. In recent years, magneto-mechano-electric (MME) energy harvesting strategy using laminated ME composites has been shown promising alternative of single-phase ME materials for harvesting low amplitude–low frequency AC magnetic field and supports the expansion of the Internet of Things (IoT) concept [5,6, 8–10]. Conventionally, magnet proof mass based cantilever structure is used as basic design for MME generator [9,10]. In contrast, the MME harvester using flexible ME nanocomposite was rarely studied. The advantage of the nanocomposites are their one step fabrication, tailor shaping, miniaturization and to the ability of largescale production and long term stability of the piezoelectric/magnetostrictive interface [11,12]. Established strategies to deploy the highest performance of the ME composites were to use inorganic/lead-based piezoelectric materials (such as, $K_{0.5}Na_{0.5}NbO_3$, PMN-PT, PZN-PT, PZT and so on) whose device configurations were mechanically stiff, brittle and sometimes heavy, with restricted biocompatibility to the human body [6,11]. Not only are those, the lead contained materials supposed to be a neurotoxin and threatening to the environment. Even, governmental policies against these kinds of hazardous substances are coming into force to research on lead-free ME composites [8]. Additionally, piezoelectric ceramics affecting at the interface regions by reactions, possess low electrical resistance and manifest high dielectric losses [12]. In contrast, ferroelectric polymers (such as, PVDF and its co-polymers with their all trans, *i.e*., TTTT conformation, named β-phase) based ME composites overcoming the aforementioned limitations and provide extra facilities of adequate flexibility, light weight, low cost, tailor size, shape, morphology with higher mechanical properties and biocompatibility in consideration of green electronics and flexible devices [11–13]. In this context, the co-polymer, polyvinylidene-co-trifluoroethylene (P(VDF–TrFE)) intrinsically possessing β-phase is a good choice of materials due to its higher dipole density in comparison to PVDF [14,15]. Trifluoroethylene residues as steric hindrance stabilizers play the key role in the copolymer, P(VDF-TrFE) to form the β-phase [14] which minimizes the need of extra physical stretching process to obtain piezoelectricity. Although, extra post-processing can improves the piezoelectric properties [15]. On the other hand, among the available magnetic materials, nickel ferrite (NiFe$_2$O$_4$/NFO) and cobalt ferrite (CoFe$_2$O$_4$/CFO) are the best known soft magnetic materials [6]. NFO is generally used as a promising magnetostrictive material due to its suitable magnetostriction, reasonable permeability, low coercivity, and high resistance which make it useful in magnetoelectric applications. Although, ME effects of PVDF/CFO and P(VDF-TrFE)/CFO composites have been extensively studied but P(VDF-TrFE)/NFO 0-3 nanocomposite has been rarely studied [12,13,16,17].

In this work, a completely rollable magneto-mechano-electric nanogenerator (MMENG) is demonstrated using P(VDF-TrFE)/NFO 0-3 type magnetoelectric nanocomposites. The piezoelectric, magnetic and magnetoelectric properties of the as fabricated nanocomposites have been systematically investigated in order to understand the structure performance correlation. The MMENG harvests the low and weak parasitic magnetic noise and operates consumer electronics. Additionally, the MMENG serves as an IoT based sensor through which the harvested signal is transmitted to a smart phone for developing a wireless position monitoring system.

## 2. Experimental section

### 2.1. NiFe$_2$O$_4$ nanoparticles synthesis

Total synthetic procedure of NiFe$_2$O$_4$ nanoparticles follows a surfactant assisted hydrothermal method. Here, Cetyltrimethylammonium bromide (CTAB, Merck India Ltd) was used as the cationic surfactant. Firstly, 0.078 M of CTAB was dissolved in 35 mL of water in order to prepare a transparent solution. Subsequently, Ferric Chloride Hexahydrate (FeCl$_3$, 6H$_2$O: Merck India Ltd) (0.092 M) and Nickel Chloride (NiCl$_2$ :Merck India Ltd) (0.045 M) were added into the solution as Fe and Ni source. Then, the solution was vigorously stirred and diluted up to 40 mL. 2 M NaOH solution and conc. ammonia solution was added drop wise into solution to adjust the pH of the solution to 11. Then, the solution was transferred into a Teflon-lined autoclave after a pre-treatment under an ultrasonic bath for 30 min. The hydrothermal was performed for 15 h at 130 °C. Finally, the nanoparticles were collected after filtration followed by annealing at 100 °C for 24 h.

### 2.2. Magnetoelectric film preparation

The ME films were prepared by heat control (~80 °C) spin coating (~ 800 rpm) of solutions of NFO nanoparticles (such as, 0.5 wt % (w/v) and 1 wt % (w/v)) added P(VDF-TrFE) (70:30, Piezotech FC30)–2 butanone (Methyl Ethyl Ketone (MEK), Merck, India) solution (16 wt% (w/v)) followed by annealing at 110 °C for 6 h. The films were named as, TrFE/NFO1 and TrFE/NFO2 for 0.5 and 1 wt % NFO added P(VDF-TrFE) films respectively and P(VDF-TrFE) where no nanoparticles were added. The thicknesses of the films were 9 μm as measured from the FE-SEM image. Finally, the ME films were electrically poled under 12 kV DC voltage and 60 °C temperature by corona poling (MILMAN).

### 2.3. Device fabrication

The magneto-mechano-electric nanogenerator (MMENG) was fabricated by pasting silver paste and attaching copper wires on both side of the ME films followed by encapsulation by polydimethylsiloxane (PDMS, Sylgard 184) layer, prepared by curing agent in 10:1 ratio and cured in an oven at 60 °C for 60m.

### 2.4. Characterization Techniques

The particle size was determined by high resolution transmission electron microscopy (HR-TEM, JEOL, JEM-2100) and confirmed by selected area electron diffraction (SAED) patterns. The X-ray diffraction (XRD) pattern was measured by X-ray diffractometer (Bruker, D8 Advance) with CuK$_\alpha$ radiation ($\lambda$=1.54 Å). The morphological study was performed by field emission scanning electron microscope (FESEM, S-4800, Hitachi). The electroactive phase change of the ME films were studied by Fourier transform infrared spectroscopy (FT-IR) (TENSOR II, Bruker) with 4 cm$^{-1}$ spectral resolution and 16 no. of scan. The $d_{33}$ values of the ME films were measured by $d_{33}$ meter (Piezotest, PM300). A Quantum Design Physical Properties Measurement System (PPMS) was used to determine the magnetic properties. The magnetoelectric measurements were carried out by well-established homemade setup, following the dynamic method [17]. The output voltage of the harvester was recorded with a digital storage oscilloscope (Agilent DSO3102A). The electric kettle used in this study was from Jaipan (JEK 35, 230 V, 50/60 Hz, 1000 W).

## 3. Results and discussion

The operational mechanism of MMENG with the TrFE/NFO 0-3 nanocomposites is depicted schematically in Fig. 1. As a matter of fact, under the AC magnetic field (Fig. 1(i)), the NFO nanoparticles of the MMENG experience magnetostriction and elongates or contracts (magneto–mechano coupling) with the frequency of the magnetic field (Fig. 1(ii)). This coupling induced strain is then transmitted to the piezoelectric layer of P(VDF-TrFE), and thereby, stress is applied ((Fig. 1(iii))) that generates electric charges. Eventually output voltage is produced across the external electrical load through direct piezoelectric effect (mechano-electric coupling). The harvested electrical power was used for powering consumer electronics (Fig. 1(iv)). In order to fabricate the ME nanocomposites, at first, magnetic NFO nanoparticles were synthesized with low temperature hydrothermal synthesis route (details are given in the experimental section). The XRD pattern with peaks indexed by (220), (311), (400), (422), (511) and (440) lattice planes confirms the formation of NFO nanoparticles with *Fd3m* cubic spinel structure (JCPDS 10-0325) (Fig. 2a) [18].

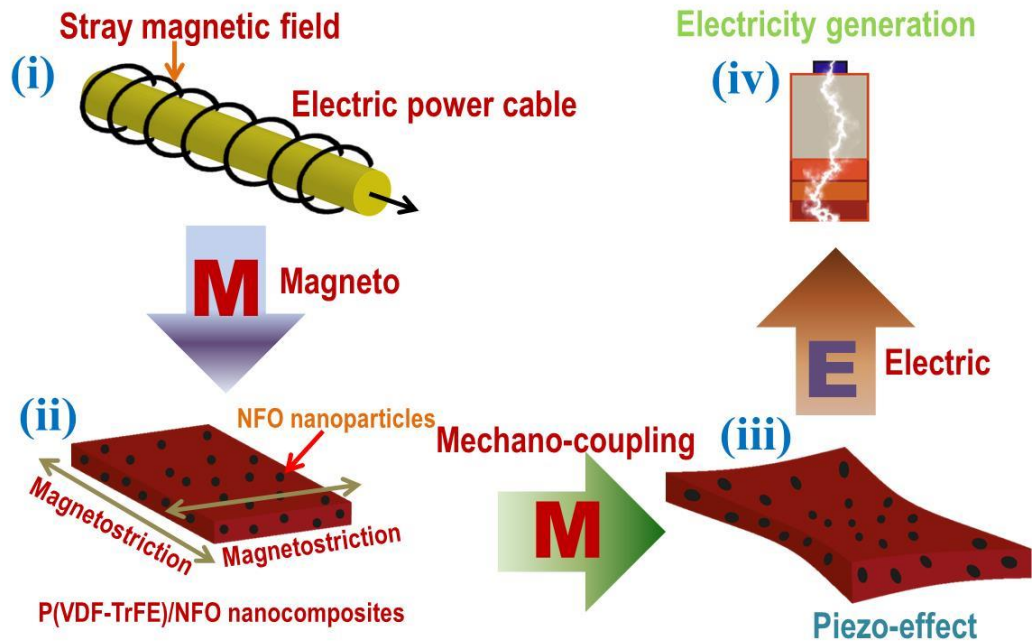

**Fig. 1.** Schematic representation of magneto-mechano-electric (MME) mechanism of TrFE/NFO 0-3 nanocomposites to harvest stray magnetic field from electric power cable.

In addition, HRTEM image shows distributed NFO nanoparticles (Fig. 2b). The inset of Fig. 2b presents histogram profile of nanoparticles diameter distribution with Gaussian fitting which reveals average diameter of 9 nm with 8 % standard deviation. The SAED pattern from a number of NFO nanoparticles, shown in Fig. 2c, further confirms face centered cubic (FCC) structure. The inset shows representative indexed lattice planes which are consistent with XRD pattern.

Fig. 2d shows the HRTEM image of few nanoparticles revealing lattice *d*-spacing of 0.29 and 0.25 nm which corresponds to (220) and (311) lattice planes respectively. The semi-transparent TrFE/NFO nanocomposite based ME films and transparent P(VDF-TrFE) film as a reference was achieved by heat control spin coating technique (Fig. 3a, details are given in the experimental section).

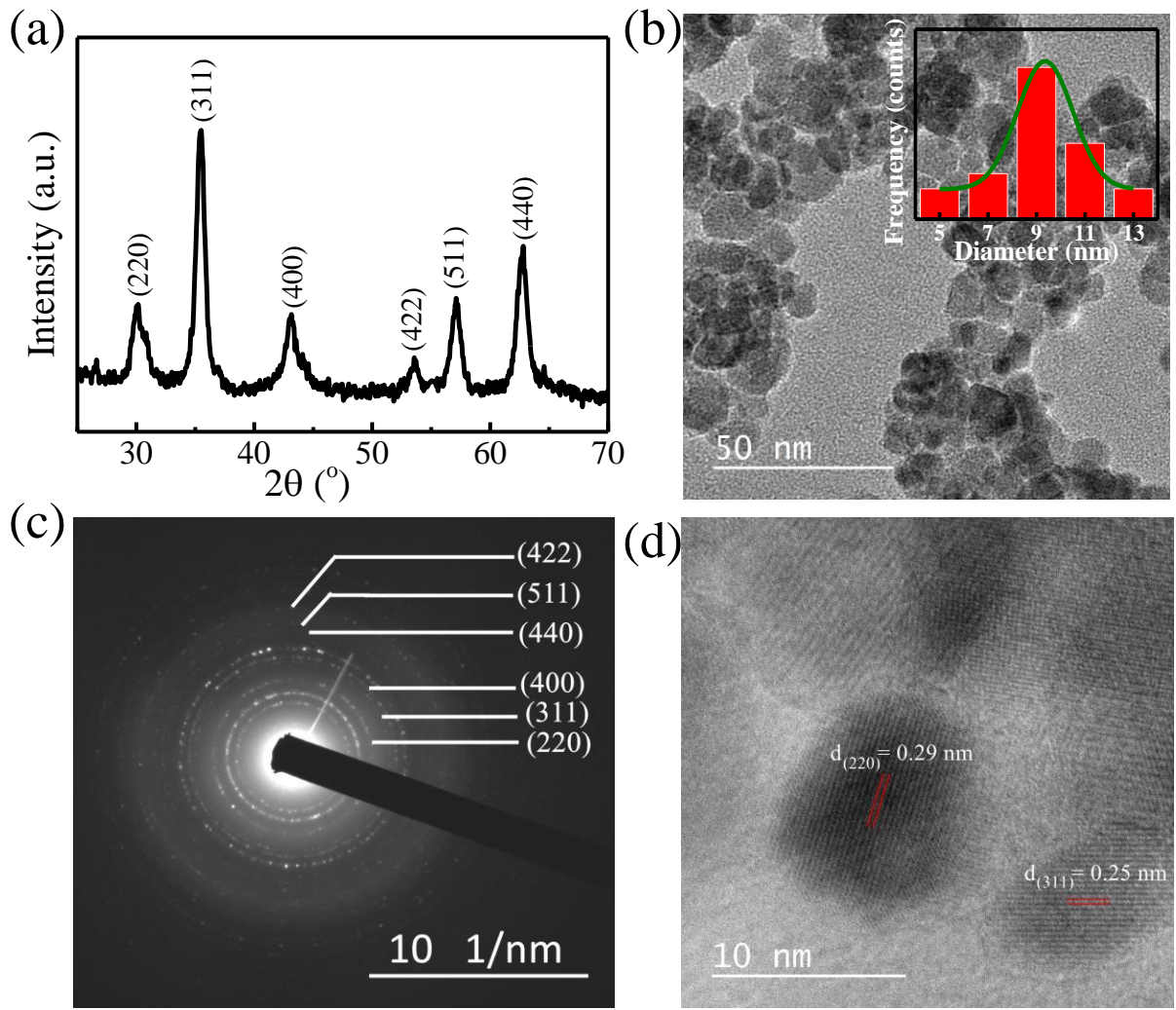

**Fig. 2.** Characterization of NFO nanoparticles. (a) XRD pattern, (b) HRTEM image with histogram profile of nanoparticles diameter distribution, (c) SAED pattern, and (d) HRTEM image of single nanoparticle with d-spacing as mentioned.

The rollable property of the TrFE/NFO nanocomposites is demonstrated in Fig. 3a. The FE-SEM images show that P(VDF-TrFE) (Fig. 3b) and TrFE/NFO nanocomposites (Fig. 3c,d) are composed of edge-on crystalline lamellae over the surface area. In addition, the NFO nanoparticles are found to be uniformly distributed over the surface area for TrFE/NFO1 (Fig. 3c) and TrFE/NFO2 (Fig. 3d) nanocomposites. The larger number of nanoparticles in TrFE/NFO2 compared to TrFE/NFO1 is clearly visible. Furthermore, the XRD pattern of the TrFE/NFO 0-3 nanocomposite films show intense peak of ferroelectric β (110/200) phase which is also present in P(VDF-TrFE) film (Fig. 4a).

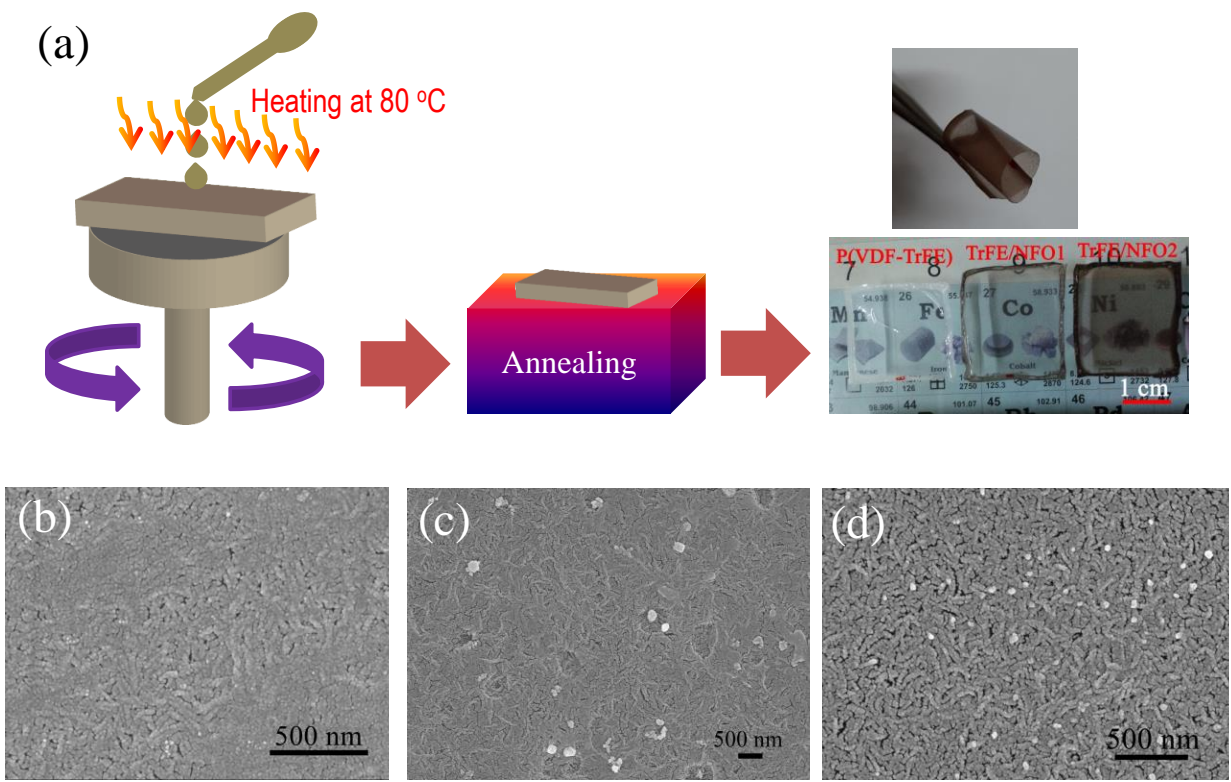

**Fig. 3.** (a) Fabrication procedure and the as fabricated flexible TrFE/NFO 0-3 nanocomposies. FE-SEM images of (c) P(VDF-TrFE), (d) TrFE/NFO1 and (e) TrFE/NFO2 films.

Note that, XRD pattern of the TrFE/NFO nanocomposites also manifests the presence of NFO nanoparticles with peaks of (311) and (400) lattice planes. From, the FTIR spectra in Fig. 4b, it is evident that TrFE/NFO2 consists of highest content of β-phase, such as, $F(\beta) \sim 84\%$, calculated from, $F_\beta = \frac{A_{841}}{\left(\frac{K_{841}}{K_{762}}\right)A_{762} + A_{841}} \times 100\%$ [19]. On the other hand, TrFE/NFO1 and P(VDF-TrFE) compose of $F(\beta) \sim 81\%$ and $F(\beta) \sim 78\%$ respectively. As a result, $d_{33}$ of the TrFE/NFO2 shows highest value of 32 pC/N in comparison to TrFE/NFO1 (~28 pC/N) and P(VDF-TrFE) (~26 pC/N) films (Fig. 4c). Apparently, the NFO nanoparticles improved the piezoelectric content and crystallinity of the polymer matrix near the interface via strong interfacial interaction with polymer chain (evident from frequency shifting in FTIR spectra, see inset of Fig. 4b) which is critically needed for good ME material [7]. Furthermore, magnetization (M) vs. magnetic field (H) loops of the TrFE/NFO 0-3 nanocomposites are measured in room-temperature and shown in Fig. 4d. The hysteresis loops show increase of magnetic moments with applied magnetic fields and the magnetization remains unsaturated at maximum applied magnetic field of 14 kOe which evidences superparamagnetic behavior of NFO in the 0-3 nanocomposites [20]. As expected, the maximum magnetic moment ($M_{max}$) of TrFE/NFO2 was higher (~ 2.5 emu g$^{-1}$) than that of TrFE/NFO1 (~ 1.6 emu g$^{-1}$) (inset of Fig. 4d) due to higher concentration of NFO nanoparticles in TrFE/NFO2. As a matter of fact TrFE/NFO 0-3 nanocomposites possess superior piezoelectric and magnetic properties. To ensure that TrFE/NFO 0-3 nanocomposites can successively convert surrounding minuscule magnetic noise into electric power, ME analysis should be performed. The ME effect of the nanocomposites were measured by the dynamic method at room temperature, which involves a superimposed condition of alternating magnetic field ($h_{ac}$) with a constant ($H_{dc}$) magnetic field (Fig. 5a) [17]. The magnetoelectric output voltage ($V_0$) and eventually magnetoelectric coupling coefficients ($\alpha = \frac{V_0}{h_{ac} t}$ where, $t$ is the thickness of the sample) were measured. The superimposed $h_{ac}$ from AC Helmholtz coils and $H_{dc}$ from a DC electromagnet were applied to the ME composites along their thickness (Z-axis) direction, as shown in Fig. 5a.

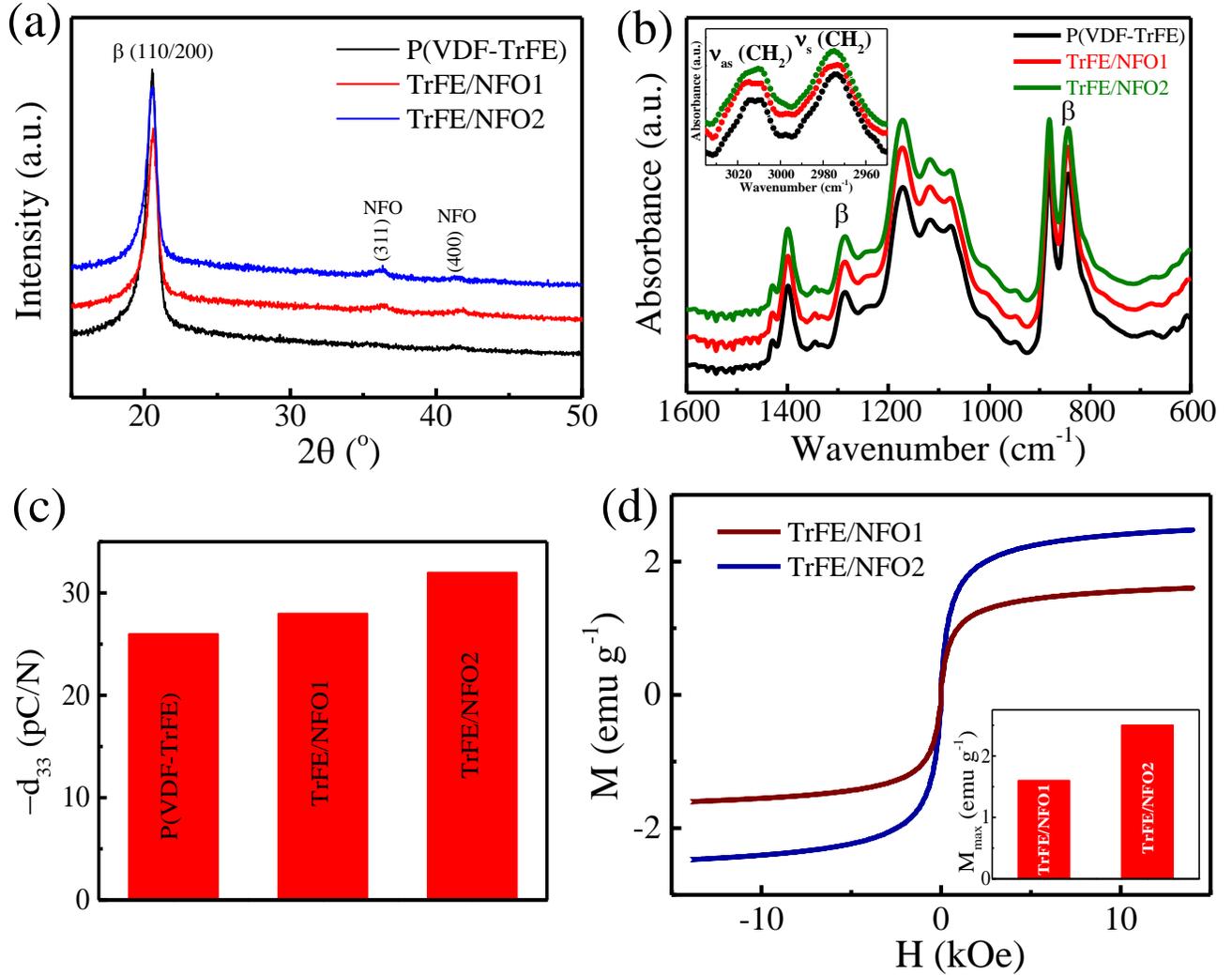

**Fig. 4.** Characterization of TrFE/NFO 0-3 nanocomposites. (a) XRD pattern, (b) FT-IR spectra, (c) measured values of $d_{33}$, and (d) magnetization (M) versus magnetic field (H) loops with maximum magnetization ($M_{max}$) of the nanocomposites in the inset.

According to previous study on PVDF/ NFO nanocomposite, lower internal field exists along the thickness direction [21] and therefore, applying magnetic field along the same direction can generate maximum magnetostrictive strain which eventually produce maximum ME output voltage. Thereby, ME output voltage, induced along the thickness (Z-axis) direction (which is also the electric poling direction) was measured using a lock-in amplifier. In order to identify the resonance frequency of the nanocomposites, $\alpha$ was measured as a function of frequency from 100–400 Hz at $h_{ac}$ ~ 20 Oe (Fig. 5b,c). The resonance frequencies were found at 170 Hz for both nanocomposites. Under this condition, the out-of-plane $\alpha$ (*i.e.*, $\alpha_{33}$) was also measured under a range of the out-of-plane DC magnetic field $H_{dc}$ ~ ± 10 kOe. At the resonance condition, the values of linear magnetoelectric coefficients were found as 5.8 mV/cm-Oe and 11.3 mV/cm-Oe for TrFE/NFO1 (Fig. 5b) and TrFE/NFO2 (Fig. 5c) respectively under the absence of $H_{dc}$ (*i.e.*, $\alpha_{33}$ at $H_{dc}$= 0 in Fig. 5d,e). When DC magnetic bias field is applied, the magnetoelectric output voltage shows a second order effect. The $\alpha_{33}$ was found to vary as a linear and anhysteretic function of $H_{dc}$ with changing polarity on changing field direction due to linear magnetostriction of the NFO (Fig. 5d,e). Interesting fact is that $\alpha_{33}$ value decreases and increases with increasing and decreasing $H_{dc}$ respectively due to negative magnetostriction effect of NFO nanoparticles [22]. Thus, the maximum value of $\alpha_{33}$ was found as 6.6 mV/cm-Oe and 11.43 mV/cm-Oe for TrFE/NFO1 (Fig. 5d) and TrFE/NFO2 (Fig. 5e) respectively. This maximum $\alpha$ value of TrFE/NFO2 is not only comparable to the theoretically predicted value of 16 mV/cm-Oe for P(VDF-TrFE)/NiFe$_2$O$_4$ 0-3 nanocomposites [23] but also larger than the P(VDF-TrFE)/Ni$_{0.5}$Zn$_{0.5}$Fe$_2$O$_4$ 0-3 nanocomposites ($\alpha_{33}$~1.35 mV/cm–Oe) [24], all-ceramic 3–3 type (Ni,Zn)Fe$_2$O$_4$/(Ba,Pb)(Zr,Ti)O$_3$ composites ($\alpha_{33}$~ 0.7 mV/cm–Oe ) [25], and many others magnetoelectric composites [12].

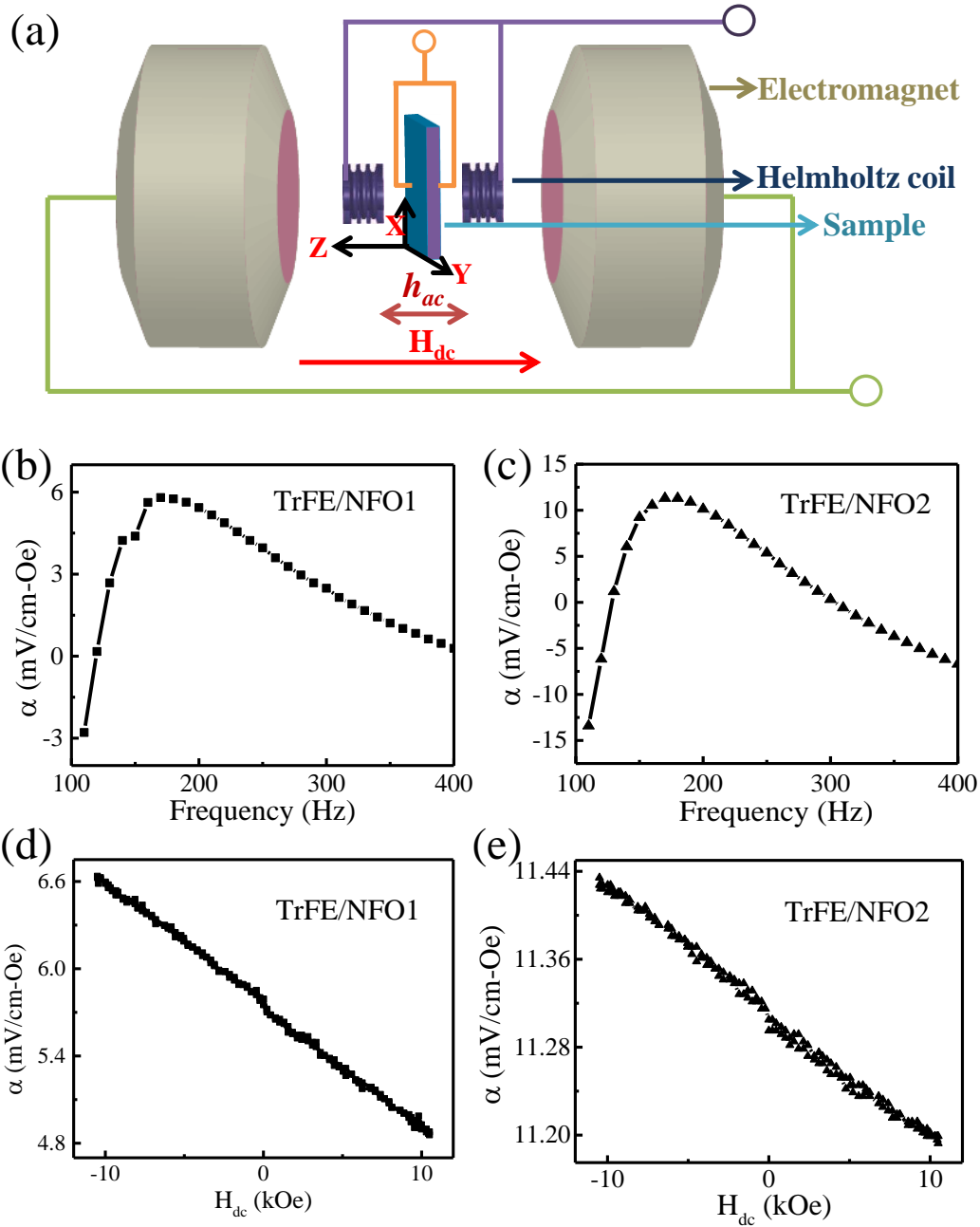

**Fig. 5.** (a) Magnetoelectric (ME) characterization set up under tunable AC and DC magnetic fields and sample orientation representation (along Z-axis) with applied magnetic field along with induced voltage direction (i.e., along Z-axis). The ME coefficients (α) of (b) TrFE/NFO1 and (c) TrFE/NFO2 as a function of AC magnetic field frequency. The ME coefficients of (d)

TrFE/NFO1 and (e) TrFE/NFO2 as a function of DC magnetic field under fixed resonance frequency of AC magnetic field.

The improved magnetoelectric effect is observed in TrFE/NFO 0-3 nanocomposites due to the strong interfacial interaction between magnetic and piezoelectric phases which are evident in FTIR spectra and FESEM images. The smaller particles size (~ 9 nm) with huge interfacial area with nano-dipoles of polymer chain favors elastic interaction with good connectivity between the piezoelectric and magnetic phases which results excellent magnetoelectric effect. It is important to note that both TrFE/NFO nanocomposites shown non-zero $\alpha$ values even at zero bias $H_{dc}$ field (Fig. 5d,e) due to self-bias ME effect which arises from the poled state of nanocomposites [26]. The TrFE/NFO1 shows 88 % and TrFE/NFO2 shows 99 % of the maximum values of $\alpha$ at zero $H_{dc}$ field. This self-bias behavior indicates that the developed ME nanocomposites based MME energy harvester can provide output electric power density even at $H_{dc} = 0$ with miniature device configuration.

The fabricated MMENG is schematically shown in Fig. 6a(i). Here, magnetic field energy harvesting has been demonstrated from stray magnetic field around the power cable of commercial home appliance of electric kettle (maximum power: 1 kW, frequency: 50/60 Hz) as shown in Fig. 6a(ii). The digital photograph of the as-fabricated MMENG is shown in Fig. 6a(iii). Additionally, the MMENG not only shown to be flexible (Fig. 6a(iv)) but also as rollable (Fig. 6a(v)). The MMENG was placed below the power cable at 0.5 mm distance to capture maximum magnetic field by the TrFE/NFO ME nanocomposites without blocking the magneto-mechano vibration.

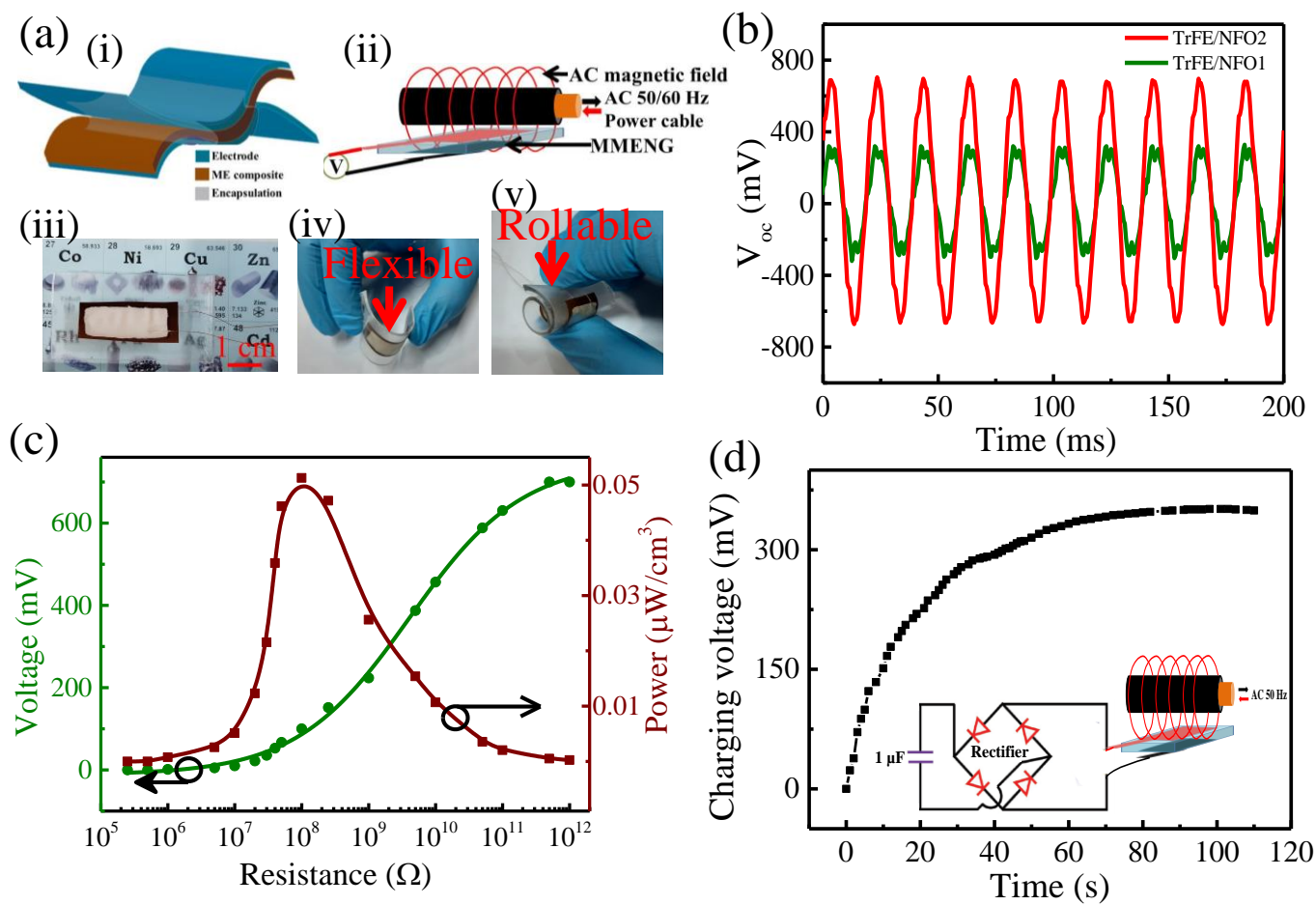

**Fig. 6**. (a) (i) Schematic of the device architecture and (ii) magnetic field energy harvesting using MMENG, (iii) digital photograph of the original device (with TrFE/NFO1 composite) which is (iv) flexible and (v) rollable, (b) harvested output voltages from TrFE/NFO based MMENGs, (c) the peak output voltage and evaluated power density as a function of load resistance, (d) capacitor charging performance of the MMENG using the circuit shown in the lower inset.

In this configuration, the AC magnetic field strength near the power cable was calculated as 1.7 mT using Ampere's law. As shown in Fig. 6b, the MMENG generated AC voltage output of sinusoidal waveform. The MMENG with TrFE/NFO1 generated peak-to-peak output voltage,

$V_{pp} \sim 640$ mV, whereas, TrFE/NFO2 generated $V_{pp} \sim 1.4$ V. Owing to the higher ME co-efficient, the voltage output from the TrFE/NFO2 was higher than that of TrFE/NFO1. Therefore, the TrFE/NFO2 based MMENG was chosen for further applications. The power output of the TrFE/NFO2 based MMENG was evaluated by measuring the peak output voltages as a function of external load resistances ($R_L$) ranging from 0.25 M$\Omega$ – 1 T$\Omega$ (Fig. 6c). The output power density (P) was evaluated using the relation, $= \frac{1}{V_s} \cdot \frac{V^2}{R_L}$, where, $V_s$ is the volume of the MMENG and V is the peak output voltage. The maximum power density was found to be 0.05 µW/cm$^3$ across the load resistance of 100 M$\Omega$. The output power from the MMENG was high enough to drive a commercial capacitor of capacitance 1 µF (Fig. 6d). The capacitor was charged up to 350 mV (0.06 µJ) within 115 s after rectifying harvested AC output using the circuit shown schematically in the inset of Fig. 6d. The energy conversion efficiency of the MMENG is expressed as, $\eta = \frac{harvested\ electric\ power}{energy\ stored\ in\ magnetic\ material} = \frac{P_{out}}{P_{in}}$. The $P_{in}$ is calculated by $P_{in} = \frac{h_{ac}^2 V_s f}{2\mu}$, where, $f$ and $\mu$ are the frequency and magnetic permeability.[5] The evaluated energy conversion efficiency thus found as 0.1 %.

Furthermore, the MMENG was demonstrated as a wireless IoT sensor. The application concept of the MMENG based position monitoring system is shown in Fig. 7a. The harvested signal by MMENG (using TrFE/NFO2) from the power cable of electric kettle was wirelessly transferred to a smart phone by Arduino microcontroller unit (MCU) based prototyping platform (Arduino UNO R3 ATMEGA 328P). The schematic of the wireless data transmission circuit is shown in Fig. 7b. The real-time digital photograph of the Arduino MCU board connected with a Bluetooth module (HC-05) and an Android-based smart phone that displays the harvested signals are shown in Fig. 7b (see Video S1 in the Supporting Information).

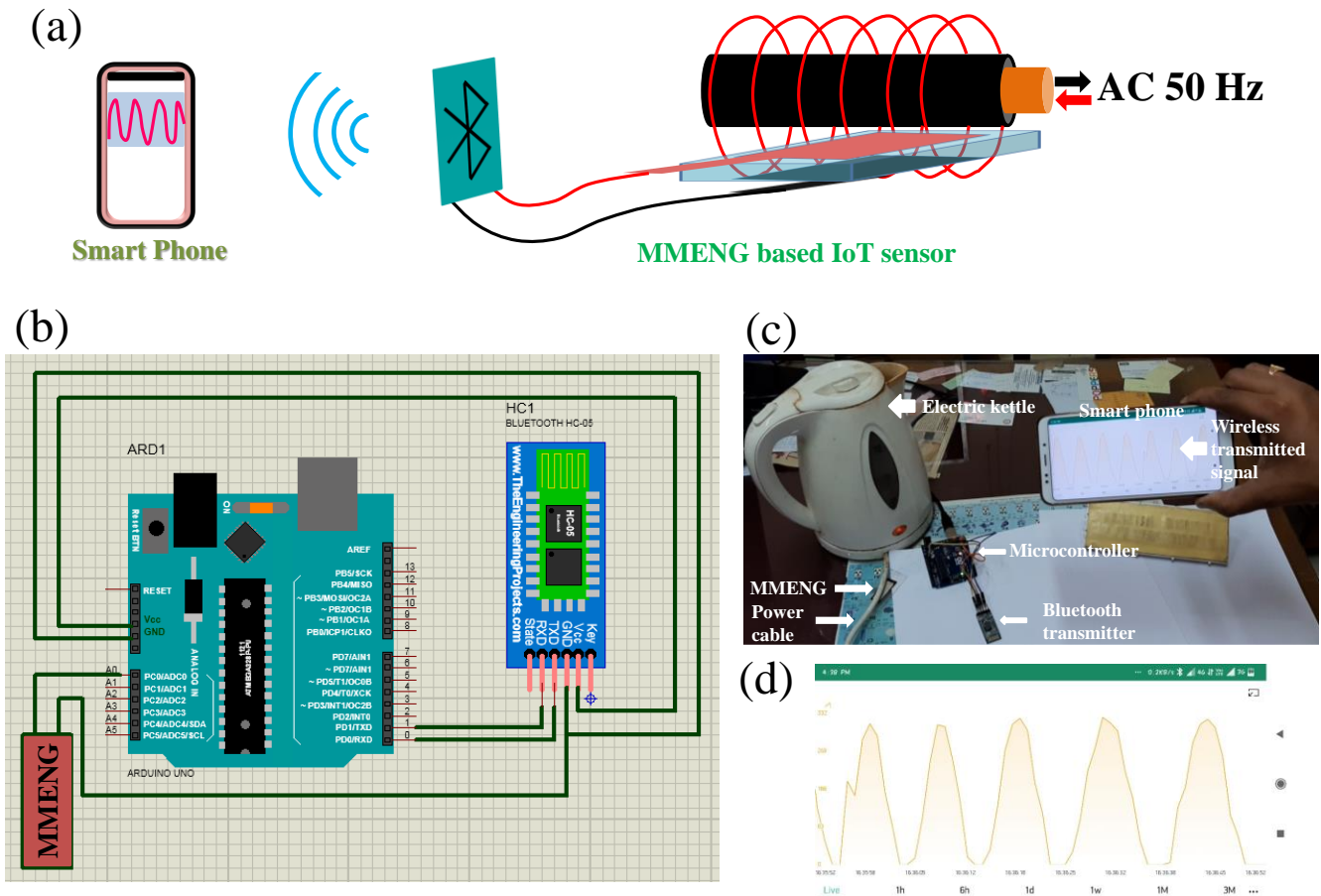

**Fig. 7.** (a) Conceptual schematics of the MMENG based position monitoring system (b) schematic diagram of the MMENG based wireless signal processing unit, (c) digital photograph of the wireless transmission of the harvested signal by MMENG from electric kettle power cable, and (d) captured image of the transmitted signal to the smart phone.

The captured wireless output signal is further shown in Fig. 7c. The MCU board converts the MME signal into 10 digital forms per second and transfers them to the smart phone via a Bluetooth transmitter. Therefore, the MMENG could be used as position monitoring IoT sensor (Fig. 7a) [27]. When the user of the IoT device such as, smart phone approaches near the

Bluetooth transmitter, the amplitude of the transmitted signal to the smart phone will be higher and when the user is far from the transmitter the amplitude will be lower in that case. Thus, the entire results ensure the feasibility of MMENG to harvest ubiquitous parasitic magnetic field and its application towards IoT based devices.

## 4. Conclusions

In summary, the energy harvesting capability of the magnetoelectric P(VDF-TrFE)/NFO 0-3 nanocomposite based MMENG was demonstrated under low-level magnetic field of 50 Hz frequency present in our living environment. The investigated piezoelectric, magnetic and magnetoelectric properties of the as fabricated nanocomposites were depended on the concentration of NFO nanoparticles. Direct magnetoelectric coupling coefficients up to 11.43 mV/cm–Oe were achieved in 10 kOe DC magnetic bias field at a resonance frequency of 170 Hz. Depending on excellent self-bias effect of the nanocomposites, the feasibility of the MMENG for real world applications was established through the power generating performance from our regular life environment and wireless data transfer through IoT based device like smart phone as a first step towards indoor wireless position monitoring system development.

**Acknowledgments**

This work was supported by the Science and Engineering Research Board (EEQ/2018/001130, SERB/1759/2014-15), Govt. of India. Krittish Roy was supported by INSPIRE fellowship IF160559. They also acknowledge instrumental facilities developed in Jadavpur University under FIST-II programme which is financially supported by DST, Govt. of India.

**Conflict of interest**

The authors declare no competing financial interests.

**Supporting Information**

Video S1 shows the wireless transfer of the harvested signal from the power cable of the electric kettle by MMENG to the smart phone.